\documentclass[twocolumn,times]{aastex62}

\usepackage{textcomp}
\usepackage{natbib}
\usepackage{epsfig,amsmath,amssymb}
\usepackage{graphicx}

\def\CII{[C\,\textsc{ii}]~158\,$\mu$m}
\def\LCII{$L_{\rm [C{\scriptscriptstyle II}]}$}
\def\LTIR{$L_{\rm TIR}$}
\def\logNHI[#1]{$\log(N_{\rm H{\scriptscriptstyle I}}/{ \rm cm^{-2}})$ = #1}
\def\kms{km~s$^{-1}$}
\def\HI{H\,\textsc{i}}

\received{2018 November 15}
\revised{2018 December 13}
\submitjournal{ApJL}

\shorttitle{\CII\ Emission from $z \sim 4$ \HI\ Absorption-Selected Galaxies} 
\shortauthors{Neeleman et al.}
 
\begin{document}
\title{[C\,\textsc{ii}]~158\,$\mu$m Emission from $z \sim 4$ \HI\ Absorption-Selected Galaxies}

\correspondingauthor{Marcel Neeleman}
\email{neeleman@mpia.de}

\author[0000-0002-9838-8191]{Marcel Neeleman}
\affiliation{Max-Planck-Institut f\"{u}r Astronomie, K\"{o}nigstuhl 17, D-69117, Heidelberg, Germany}
\author[0000-0002-9757-7206]{Nissim Kanekar}
\altaffiliation{Swarnajayanti Fellow}
\affiliation{National Centre for Radio Astrophysics, Tata Institute of Fundamental Research, Pune University, Pune 411007, India}
\author[0000-0002-7738-6875]{J. Xavier Prochaska}
\affiliation{Department of Astronomy \& Astrophysics, UCO/Lick Observatory, University of California, 1156 High Street, Santa Cruz, CA 95064, USA}
\affiliation{Kavli Institute for the Physics and Mathematics of the Universe (Kavli IPMU)}
\author[0000-0002-9946-4731]{Marc A. Rafelski}
\affiliation{Space Telescope Science Institute, Baltimore, MD 21218, USA}
\affiliation{Department of Physics \& Astronomy, Johns Hopkins University, Baltimore, MD 21218, USA}
\author[0000-0001-6647-3861]{Chris L. Carilli}
\affiliation{National Radio Astronomy Observatory, Socorro, NM 87801, USA}
\affiliation{Cavendish Astrophysics Group, University of Cambridge, Cambridge CB3 0HE, UK}
 
\begin{abstract}
We report on a search for the \CII\ emission line from galaxies associated with four high-metallicity damped Ly$\alpha$ absorbers (DLAs) at $z \sim 4$ using the Atacama Large Millimeter/sub-millimeter Array (ALMA). We detect \CII\ emission from galaxies at the DLA redshift in three fields, with one field showing two \CII\ emitters. 
Combined with previous results, we now have detected \CII\ emission from 
five of six galaxies associated with targeted high-metallicity DLAs 
at $z \sim 4$. 
The galaxies have relatively large impact parameters, $\approx 16 - 45$~kpc, \CII\  line luminosities of $(0.36 - 30) \times 10^8$~L$_\odot$, and rest-frame far-infrared properties similar to those of luminous Lyman-break galaxies, with star-formation rates of $\approx 7 -110$~M$_\odot$~yr$^{-1}$. Comparing the absorption and emission line profiles yields a remarkable agreement between the line centroids, indicating that the DLA traces gas at velocities similar to that of the \CII\ emission. This disfavors a scenario where the DLA arises from gas in a companion galaxy. These observations highlight ALMA's unique ability to uncover a high redshift galaxy population that has largely eluded detection for decades.
\end{abstract}


\section{Introduction}
Neutral atomic gas plays a central role in the formation and evolution of galaxies. Although simulations based on a $\Lambda$-cold dark matter ($\Lambda$-CDM) cosmology suggest that gas falling onto dark matter halos from the cosmic web is predominantly ionized, as the accreted gas migrates inward, the increase in density and pressure causes most of the gas to become neutral \citep[e.g.,][]{Keres2005, Dekel2009}. The neutral gas can then cool further to form molecular gas, which eventually forms stars. Neutral atomic gas thus acts as a reservoir of fuel for future star formation; studying its properties, and its evolution with redshift, allows us to understand how galaxies assemble their baryonic mass and convert this into stars.

\begin{figure*}
\includegraphics[width=\textwidth]{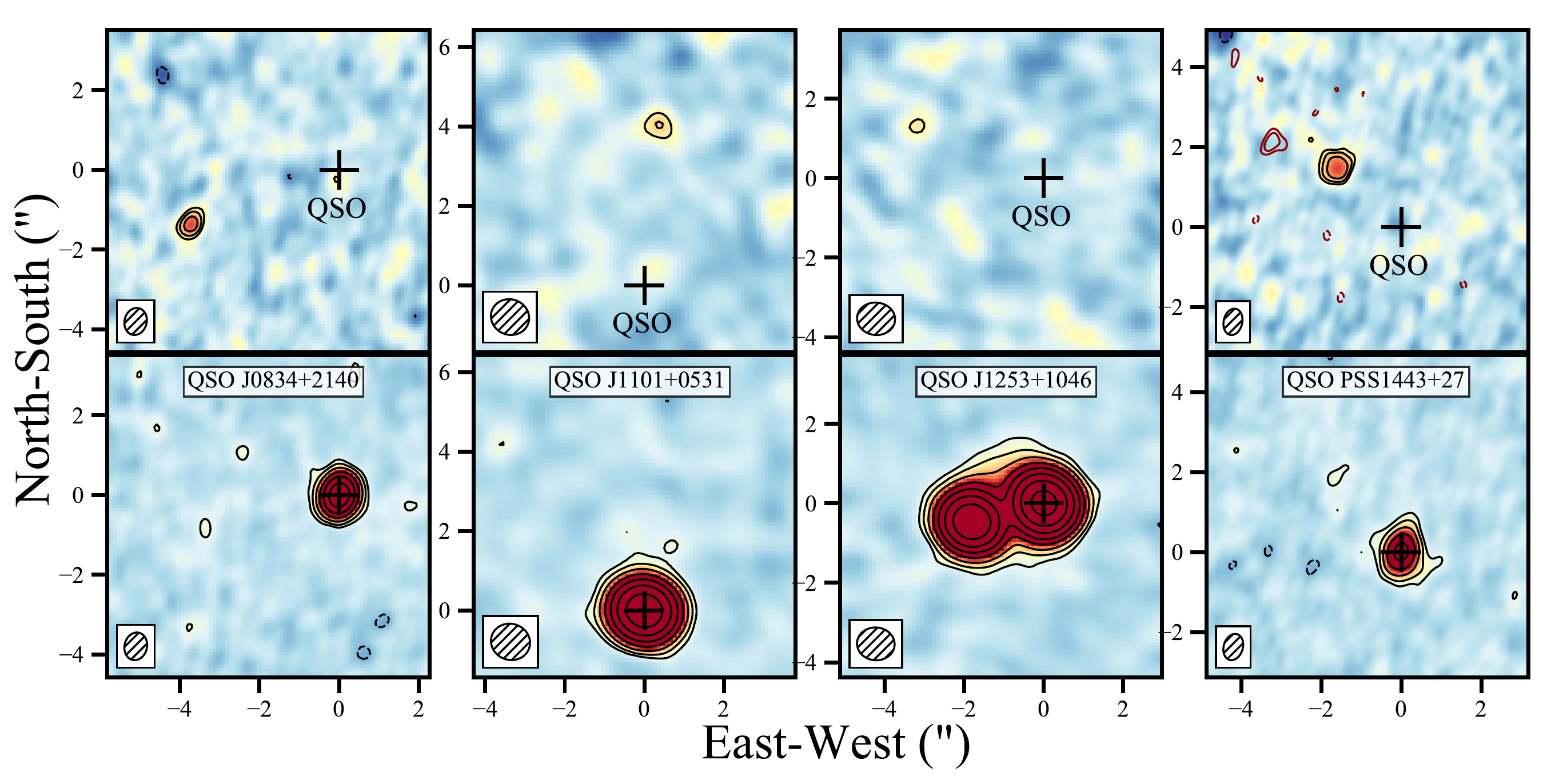}
\caption{\emph{Top panels:} Integrated \CII\ flux density maps over channels containing emission (see Fig. \ref{fig:Abs_CIISpec}). For the sole \CII\ non-detection (DLA~J1253$+$1046), the line is integrated over the central 100~\kms\ around the DLA redshift. The red contours in the rightmost panel (QSO~PSS1443$+$27) mark a second \CII\ emitter offset by $\approx 300$ \kms\ from the other emission. Contours start at $3\sigma$, with successive contours increasing by $\sqrt{2}$. \emph{Bottom panels:} Continuum images at $\sim 350$~GHz of the four quasar fields. The plus sign marks the position of the peak of the continuum emission from the quasar. Contours are drawn at [3,6,12,...]$\times \sigma$. Negative contours are dashed. The ALMA synthesized beam is shown in the bottom left corner.}
\label{fig:Cont_Mom0}
\end{figure*}

Unfortunately, it is not possible today to directly detect \HI\ emission from galaxies at $z \gtrsim 0.5$ in reasonable integration times \citep{Kanekar2016,Fernandez2016}. The Ly$\alpha$ absorption signature provides the only opportunity to detect this gas phase at high redshifts, during the epoch of galaxy assembly. The strongest Ly$\alpha$ absorbers, the so-called damped Ly$\alpha$ absorbers \citep[DLAs;][]{Wolfe2005}, trace the bulk of the neutral atomic gas in the universe at all redshifts \citep[$\approx 80\%$; e.g.,][]{Omeara2007,Noterdaeme2012}. Studying DLAs therefore allows us to directly probe the evolution of \HI. However, to study the interplay between these \HI\ reservoirs and galaxies, we need to characterize the galaxies with which the DLAs are associated.

Detecting the galaxies associated with high-$z$ DLAs is challenging \citep[e.g.,][]{Kulkarni2006,Fumagalli2015}. This has previously been assumed to be due to the inherent faintness of the DLA galaxies compared to the bright background quasar. Today, there are $\approx 20$ DLAs at $z \sim 2-3$ with detections of their associated galaxies at optical/near-infrared wavelengths \citep[e.g.,][]{Krogager2017}. Complementary to this approach, \citet{Neeleman2016b} showed that the galaxies associated with Ly$\alpha$ absorbers can be detected at millimeter wavelengths, in their CO emission with the Atacama Millimeter/sub-millimeter Array (ALMA). Subsequent ALMA studies have yielded a high success rate in CO detections from galaxies associated with DLAs out to $z \approx 2.5$ \citep{Moller2018,Kanekar2018,Neeleman2018,Fynbo2018, Klitsch2019}, and possibly beyond \citep{Dodorico2018}. 

For galaxies at even higher redshifts, the fine structure line of singly-ionized carbon, \CII, shifts into the ALMA observing bands. In \citet{Neeleman2017}, we showed that the \CII\ line can be used to identify galaxies associated with DLAs at $z \sim 4$. In this Letter, we report on an ALMA search for \CII\ emission in four additional DLAs at $z \gtrsim 4$. We assume a standard, flat, $\Lambda$-CDM cosmology with $\Omega_\lambda = 0.7$ and $H_0 = 70~\text{km}~\text{s}^{-1}~\text{Mpc}^{-1}$.

\begin{figure*}
\includegraphics[width=\textwidth]{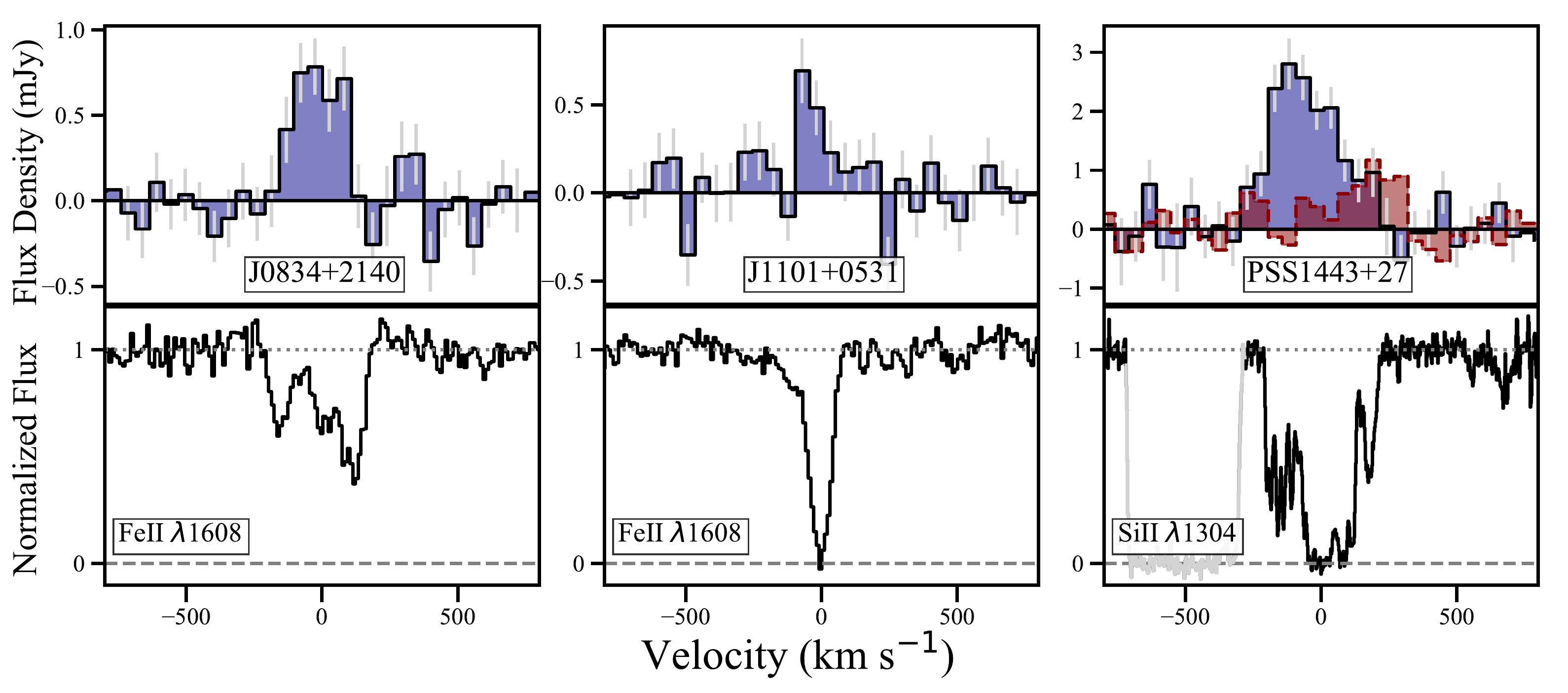}
\caption{Comparison between the \CII\ emission line profiles (\emph{top panels}) obtained from the region with velocity-integrated \CII\ emission of $\geq 2\sigma$ significance (Fig.~\ref{fig:Cont_Mom0}), with a low-ionization metal-line absorption profile (transition indicated at the bottom right; \emph{bottom panels}). The $1\sigma$ error on the \CII\ flux densities is shown in gray. Red in the rightmost panel marks the \CII\ emission line profile of the second emitter in this field (see text and Fig. \ref{fig:Cont_Mom0}). This figure exemplifies the remarkable agreement between the centroids of the metal-line absorption and the \CII\ emission profiles.}
\label{fig:Abs_CIISpec}
\end{figure*}

\section{Sample Selection, Observations and Reduction}
We used the ALMA Band-7 receivers to search for \CII\ emission in the fields of four DLAs at $z \sim 4$ between UT 2016 March 16 and UT 2017 January 29 (ALMA proposal ID's: 2015.1.01564.S and 2016.1.00569.S; PI: Neeleman). The four targets were selected from a parent sample of DLAs for which high-resolution ($R > 10\,000$) optical spectra are available, allowing an accurate determination of the gas metallicity \citep{Rafelski2012,Rafelski2014}. All selected DLAs have relatively high metallicities, [M/H]~$\geq -1.36$, which is higher than the median DLA metallicity at $z \sim 4$ \citep[$\text{[M/H]}~\approx -1.8$;][]{Rafelski2014}. The high metallicity is an indicator that these DLAs are associated with more massive galaxies \citep{Moller2013,Neeleman2013,Christensen2014}, and therefore relatively high star formation rates (SFRs). The correlation between \CII\ line luminosity and SFR in the local Universe \citep[e.g.,][]{DeLooze2014} further indicates that such high-metallicity DLAs are the best candidates for detections of \CII\ emission.

The ALMA observations used four 1.875~GHz bands, each sub-divided into 128 channels. For each target, one of the bands was centered on the expected redshifted \CII\ line frequency, while the remaining three bands were used to obtain a continuum image of the field. The on-source times were $1.2 - 2.8$~hours. All of the data were calibrated using the ALMA pipeline, in the Common Astronomy Software Applications \citep[CASA;][]{McMullin2007} package. Additional data editing was performed in CASA after the initial calibration. Two of the targets, QSO~J1101$+$0531 and QSO~J1253$+$1046, were sufficiently bright to perform self-calibration, which was done in the Astronomical Image Processing System \citep[AIPS;][]{Greisen2003} package. The final spectral cubes and continuum images were obtained using natural weighting, to maximize sensitivity, using the task \texttt{tclean} in CASA. Details of the sources and observations are provided in Table~\ref{tab:obs}, and the integrated \CII\ flux density and continuum images of the four fields are shown in Fig.~\ref{fig:Cont_Mom0}.

\section{Results}
\subsection{DLA~J0834$+$2140}
The $z = 4.3900$ DLA towards QSO~J0834$+$2140 has an \HI\ column density of \logNHI[21.0 $\pm$ 0.2] and a metallicity of $-1.30 \pm 0.20$. The multi-component velocity profile spans a large velocity interval, with a velocity spread $\Delta V_{90} \approx 290$~\kms\ \citep[e.g.,][]{Prochaska1997}. This is significantly larger than the median $\Delta V_{90}$ in DLAs, $\approx 72$~\kms\ \citep{Neeleman2013}, and indicates that the DLA is tracing either neutral gas belonging to a massive galaxy or a complex of lower-mass galaxies \citep[e.g.,][]{Bird2015}. Many metal line species are seen in the high resolution absorption spectrum of the DLA, including C$^+$ \citep{Rafelski2012}, evidence that the gas has been enriched.

Our ALMA data yield a clear ($8.7\sigma$) detection of an emission line at an impact parameter of 4\farcs0 southeast of the quasar (Fig. \ref{fig:Cont_Mom0}), which we identify as the \CII\ emission line at a redshift of $z = 4.3896$. At this redshift, an impact parameter of 4.0$''$ corresponds to 27~kpc. The total velocity-integrated line flux density is $0.173 \pm 0.020$~Jy~\kms, implying a \CII\ line luminosity, \LCII, of $(1.02 \pm 0.12) \times 10^8$~L$_\odot$ \citep[e.g.,][]{Solomon1992}. Within the positional uncertainty of the ALMA observations, a $3.7\sigma$ continuum feature is also seen at the position of the \CII\ emission with a flux density of $70 \pm 22$~$\mu$Jy. Besides the quasar, this is the most significant emission of the continuum image. Therefore, we tentatively take this as the continuum flux measurement of the \CII\ emitter. The total infrared luminosity, \LTIR, as determined from a modified black-body fit to the continuum measurement is $(6.6 \pm 2.1) \times 10^{10}$~L$_\odot$. We note that the quoted uncertainties on \LTIR\ are observational, and there is an additional systematic uncertainty of $\approx 0.5$~dex when estimating \LTIR\ from a measurement at a single wavelength \citep{Neeleman2017}. Using the conversion rate between SFR and 160~$\mu$m continuum emission \citep{Calzetti2010}, we obtain an SFR of $(7 \pm 2)$~M$_\odot$~yr$^{-1}$ for this galaxy.

\subsection{DLA~J1101$+$0531}
The $z = 4.3446$ DLA towards QSO~J1101$+$0531 has an \HI\ column density of \logNHI[21.3 $\pm$ 0.1] and a metallicity of $-1.07 \pm 0.12$. The low-ionization metal lines show a narrow velocity spread, $\Delta V_{90} = 60$~\kms. This is lower than expected from the correlation between metallicity and $\Delta V_{90}$ \citep{Neeleman2013}, and may suggest that the DLA traces metal-rich gas around a low-mass galaxy or that the sightline intersects relatively unperturbed gas around a more massive system. 

The ALMA observations show a $5.2\sigma$ emission feature 4\farcs0 north of the quasar (Fig. \ref{fig:Cont_Mom0}). The redshift of this \CII\ emission line is 4.3433, which is within 75 \kms\ of the redshift of the DLA, and the velocity-integrated line flux density is $(0.12 \pm 0.04)$~Jy~\kms\ yielding \LCII\ = $(7.0 \pm 2.3) \times 10^7$~L$_\odot$. No continuum is detected at the position of the \CII\ emission, putting an upper limit on the \LTIR\ and SFR of $< 7.0 \times 10^{10}$~L$_\odot$ and $< 7$~M$_\odot$~yr$^{-1}$, respectively.

\subsection{DLA~J1253$+$1046}
The $z=4.6001$ DLA towards QSO~J1253$+$1046 is both our highest-redshift and our lowest-metallicity target ([M/H] = $-1.36 \pm 0.16$). Note that the DLA metallicity is still above the average DLA metallicity at $z \sim 4$. Its \HI\ column density and $\Delta V_{90}$ are both low, \logNHI[20.3 $\pm$ 0.15] and $\Delta V_{90} \approx 70$~\kms\, perhaps indicating that the DLA is probing a less-massive galactic halo compared to our other targets \citep{Neeleman2013}.

We find no evidence of significant ($\gtrsim 5\sigma$) line emission in our ALMA images, near the DLA redshift. However, the 339~GHz continuum image shows a bright source 1.9$''$ southeast of the quasar. No emission lines are detected from this source in any of our four ALMA bands, consistent with a scenario in which the source is at the quasar redshift (which was not covered in the \CII\ line). Assuming that the source is indeed at the quasar redshift ($z = 4.908$), the measured continuum flux density of $(9.9 \pm 0.6)$~mJy yields \LTIR=$1 \times 10^{13}$~L$_{\odot}$ and a SFR of $\approx1000$~M$_\odot$~yr$^{-1}$, typical of high-$z$ sub-millimeter galaxies (SMGs) \citep{Carilli2013a}. If the SMG is confirmed to be at the quasar redshift, the system would be similar to the well-studied quasar-SMG pair, QSO~BR1202$-$0725 at $z \approx 4.7$ \citep[e.g.,][]{Omont1996, Carilli2013b}, except that the separation between SMG and QSO is smaller here, $\approx 12$~kpc. We stress that this galaxy is \emph{not} associated with the DLA, as the expected \CII\ line luminosity of the SMG is significantly above the detection limit of our ALMA observations.

\begin{figure}
\includegraphics[width=0.47\textwidth]{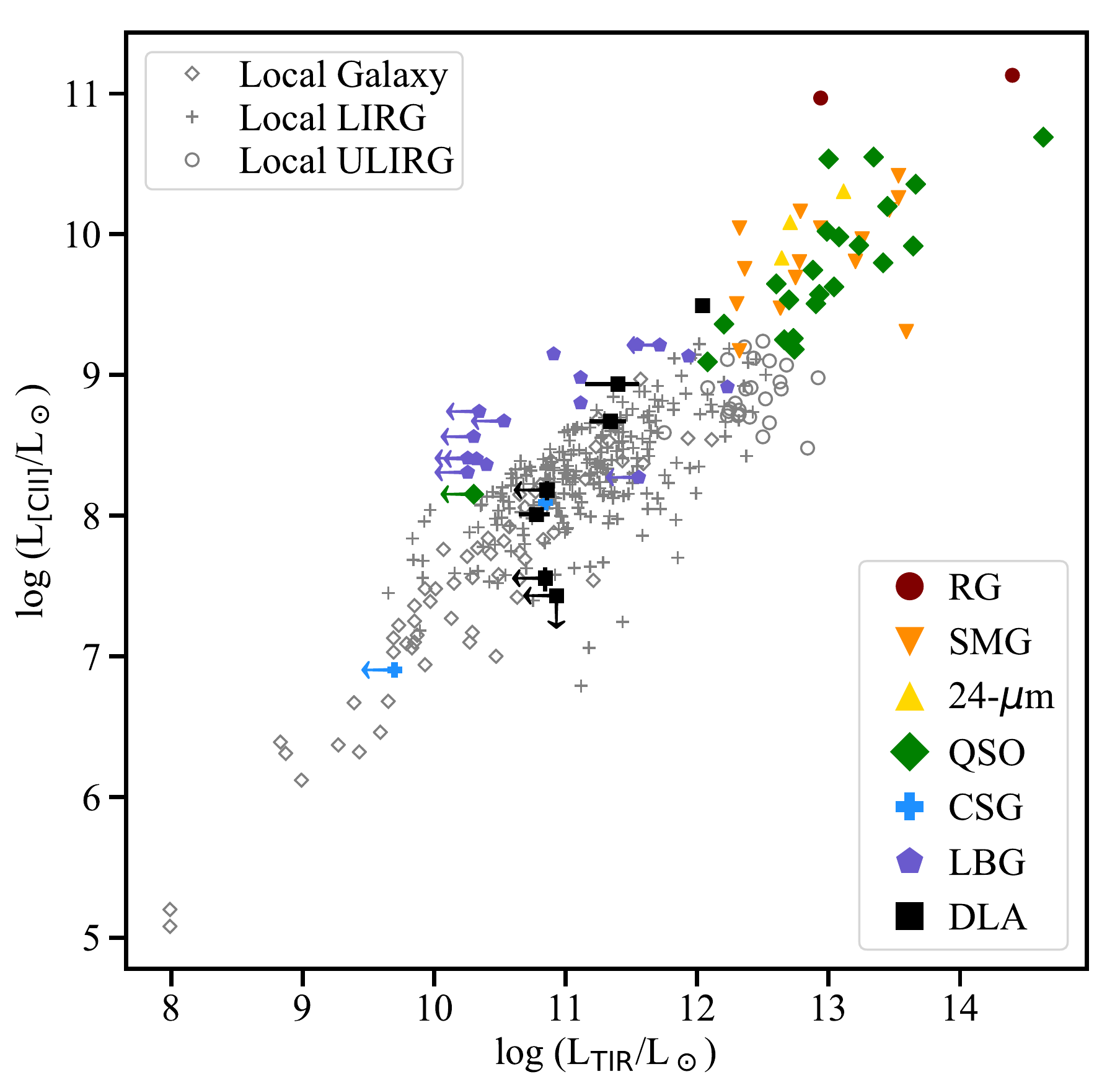}
\caption{ \CII\ line luminosity plotted against the total FIR luminosity for a sample of low-redshift galaxies (gray symbols) and a sample of high-redshift galaxies (color-coded by galaxy type). The galaxies associated with DLAs at $z \approx 4$ discussed in this Letter are shown as black squares, and follow approximately the same scaling relationship as local luminous infrared galaxies, and high-$z$ LBGs. Typical uncertainties on the high-redshift  measurements are driven by the systematic uncertainty in the determination of \LTIR, and is $\approx$0.5 dex.}
\label{fig:LCIILFIR}
\end{figure}

\subsection{DLA~PSS1443$+$2724}
The $z=4.2241$ DLA towards PSS1443$+$2724 has \logNHI[21.3 $\pm$ 0.1]\ and a metallicity of [M/H]$= -0.95 \pm 0.20$ \citep{Prochaska2001}. Strong low-ion absorption is seen over a wide range of velocities, with $\Delta V_{90} = 284$~\kms. In simulations, such large velocity spreads are typically seen in lines of sight passing through multiple halos \citep{Bird2015}. No optical counterpart was identified in deep ground-based imaging, with a $3\sigma$ $R$-band limit of $M_{\rm AB} = 26.9$ \citep{Prochaska2002}.

Our ALMA data reveal strong ($>10\sigma$) line emission at an impact parameter of 2\farcs3 northeast of the quasar. We identify this as redshifted \CII\ line emission from a $z = 4.2256$ galaxy, only $-84$~\kms\ offset from the DLA redshift. The velocity-integrated line flux density of $(0.846 \pm 0.063)$~Jy~\kms\ yields a \CII\ line luminosity of $(4.7 \pm 0.4) \times 10^8$~L$_\odot$. Additionally, at the position of the \CII\ emitter, a $3.8\sigma$ excess is observed in the 356~GHz continuum. As in J0834$+$2140, besides the QSO, this is the most significant continuum emission. We therefore assume this emission arises from the \CII\ emitter. The total flux density of $(161 \pm 44)$~$\mu$m yields \LTIR=$(14 \pm 4) \times 10^{10}$~L$_\odot$, and an SFR = $(15 \pm 4)$~M$_\odot$~yr$^{-1}$. This is the only \CII\ emitter of our present sample that is resolved, with a size of ($0\farcs58 \pm 0\farcs17 \times 0\farcs4 \pm 0\farcs3$), corresponding to an emission region of only a few kpc. 

In addition to the above \CII\ line emission, the ALMA cube shows a second line feature, close to the DLA redshift. This $6.4\sigma$ feature is detected at an impact parameter of 4\farcs0 northeast of the quasar, which is 1\farcs5 distance from the other line emission. Assuming that this corresponds to redshifted \CII\ emission, the emission velocity is offset by $+200$~\kms\ from the DLA redshift. This is well within the velocity spread of the absorption (see Fig.~\ref{fig:Abs_CIISpec}). The velocity-integrated line flux density is $(0.274 \pm 0.043)$~Jy~\kms, yielding \LCII=$(1.5 \pm 0.2) \times 10^8$~L$_\odot$. No continuum emission is detected at the position of the \CII\ emission, yielding a $3\sigma$ SFR limit of 7~M$_\odot$~yr$^{-1}$.

\section{Discussion and Summary}
\subsection{\CII\ Emission from Absorption-Selected Galaxies}
The primary aim of the ALMA observations was to use \CII\ emission to identify the galaxies associated with DLAs at $z \sim 4$. Together with the results of \citet{Neeleman2017}, the success rate of this program has been remarkable, with 5 detections out of 6 targets. Fig.~\ref{fig:LCIILFIR} shows the total far-infrared (FIR) luminosity plotted against the \CII\ line luminosity for the galaxies of our sample, along with a set of low-redshift and high-redshift galaxy populations. The figure indicates that the FIR properties of galaxies selected through absorption are similar to those of luminous Lyman-break galaxies (LBGs) and color-selected galaxies (CSGs), supporting earlier conclusions that at least the most metal-enriched DLAs are associated with galaxies similar to luminous LBGs \citep{Moller2002,Neeleman2017}. Selecting galaxies through absorption, therefore provides a unique, complementary approach to study `normal' galaxies at high redshift.

\begin{figure}
\includegraphics[width=0.47\textwidth]{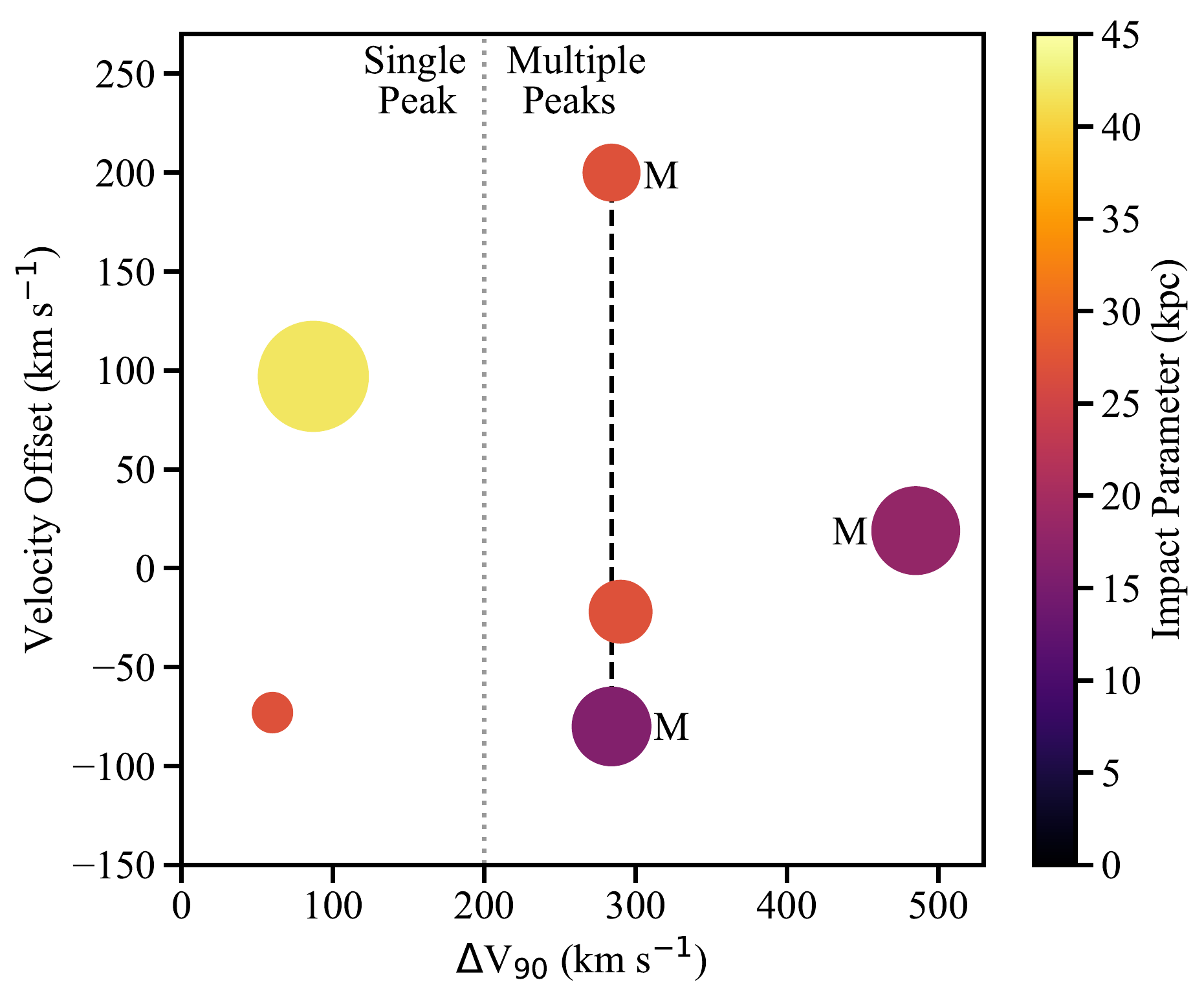}
\caption{Velocity offset between the centroid of the \CII\ emission and the low-ionization metal line absorption as a function of absorption velocity spread, $\Delta V_{90}$. The color of the points indicates the impact parameter between the \CII\ emitter and the DLA, while the size of the points is proportional to the velocity spread of the \CII\ emission. No significant correlation is found between any of these parameters in our (small) sample. However, we do find that DLAs with very large $\Delta V_{90}$ ($\gtrsim 200$~\kms) predominantly arise in systems with multiple density peaks (i.e., galaxies), consistent with predictions from simulations \citep{Bird2015}. The pair of \CII\ emitters in the field of DLA~PSS1443$+$2724 are indicated by the dashed vertical line, whereas those DLAs that are part of either an active merger or have multiple galaxies detected at their redshift are marked with an `M'.}
\label{fig:VComp}
\end{figure}

\subsection{Sizes and Impact Parameters of \CII\ Emitters}
One of the most interesting results of our ALMA studies is the relatively large impact parameter ($\approx 16-45$~kpc) between the absorbers and the \CII\ emitters \citep[including the two systems of][]{Neeleman2017}. This indicates that the neutral gas probed by high-redshift, high-metallicity DLAs \emph{is not directly responsible for star formation in the associated galaxies, but probes a more extended \HI\ envelope around these galaxies}. While \HI\ in nearby galaxies is also typically far more extended than the stars \citep[e.g.,][]{Briggs1980}, the high \HI\ column densities ($\gtrsim 10^{21}$~cm$^{-2}$) of our DLAs suggests that \HI\ reservoirs around high-$z$ galaxies are more clumpy, since such high \HI\ column densities are rarely seen at these large radial distances in nearby galaxies \citep[e.g.,][]{Zwaan2005}. In addition, the high gas metallicity ($\sim 1/10^{\rm th}$ solar) indicates that metals are efficiently mixed with the gas and transported to these distances.

Further, the large impact parameters imply that the associated galaxies are outside the point spread function (PSF) of the quasar for most modern optical/near-infrared instruments. Previous searches for high-$z$ DLA hosts have often attributed non-detections to the possibility that the galaxy might lie below the quasar PSF \citep[e.g.,][]{Kulkarni2006}. None of our $z \approx 4$ galaxies would have been below the quasar PSF for typical imaging studies. This is in agreement with the non-detection of star formation at the location of the DLA for a study where the quasar PSF was not an issue \citep[][]{Fumagalli2015}. Our results thus indicate that at least for high-metallicity, high-redshift DLAs, the quasar PSF may not play a significant role in obscuring galaxies associated with DLAs. The non-detection of these systems in typical optical imaging may instead be explained by significant dust obscuration of the galaxy. ALMA CO and \CII\ studies of DLA host galaxies thus provide a unique, complementary view to traditional optical imaging. We note, in passing, that some of these galaxies would not have been detected by the triple long-slit experiment \citep{Fynbo2010,Krogager2017}, due to their large angular separation (even in the absence of significant dust obscuration).

\subsection{Kinematics of Emission and Absorption Lines}
Fig.~\ref{fig:Abs_CIISpec} shows the \CII\ emission line profile compared with a low-ionization metal absorption line profile from the DLA. There is a striking agreement between the DLA redshift and the redshift of the \CII\ emission in the figure \citep[and in the corresponding figure of][]{Neeleman2017}. In all of the 5 detected \CII\ emitters, the velocity difference between the emission and absorption line centroids is $< 100$~\kms. Only the second, fainter and more distant, \CII\ emitter towards PSS~1443$+$2724, has a velocity offset of $+200$~\kms\ from the centroid of the absorption line. This striking agreement disfavors a scenario whereby the DLA is probing gas solely associated with another, fainter ---and presumably closer--- galaxy, as one would then expect to see, on average, a larger velocity offset between the DLA absorption and the emission from the \emph{unrelated} galaxy detected here \citep[see also][]{Neeleman2018,Fynbo2018}. To be specific, if we assume the bright \CII\ emitter resides in a moderate halo mass of $10^{11.5}$~M$_\odot$, then the expected virial velocity is $\approx 350$~\kms, which ---accounting for projection effects--- results in typical velocity offsets of $\approx 200$~\kms.

The distribution of $\Delta V_{90}$ values in DLAs has long been known to be skewed towards large values \citep{Prochaska1997}, implying either large massive galaxies, or that the sightline crosses multiple smaller density structures/peaks \citep[e.g.,][]{Prochaska1997,Bird2015}. Fig.~7 of \citet{Bird2015} shows that the fraction of DLAs arising from multiple density peaks in simulations rises sharply for $\Delta V_{90} \gtrsim 200$~\kms. Interestingly, two of the three systems in our sample with $\Delta V_{90} > 200$~\kms, show evidence for multiple \CII\ emitters: DLA~J1201$+$2117 appears to have two distinct \CII\ components in the process of merging \citep{Neeleman2017}, whereas DLA~PSS1443$+$2724 shows two clearly distinct (both spatially and spectrally) \CII\ emitters. Our ALMA observations thus corroborate the hypothesis that DLAs with large $\Delta V_{90}$ values, $\gtrsim 200$~\kms, are likely to arise from sightlines that intersect multiple density peaks (see Fig.~\ref{fig:VComp}).

\subsection{Concluding Remarks}
ALMA searches for \CII\ emission from galaxies associated with high-metallicity DLAs are proving to be an efficient way to identify and study this hitherto-elusive high-$z$ galaxy population, allowing us to detect \CII\ emission in galaxies with SFRs as low as 7 M$_\odot$~yr$^{-1}$. Our ALMA observations \citep[including the two systems of][]{Neeleman2017} have identified the galaxies associated with five DLAs at $z \approx 3.8-4.4$ via their redshifted \CII\ emission. The inferred \CII\ line and FIR continuum luminosities are consistent with the DLA galaxies being similar to luminous LBGs at these redshifts. While this correspondence may not be surprising in hindsight, it may pose a serious challenge to the current paradigm of galaxy formation which predicts that the majority of galaxies at $z \sim 4$ have low SFRs and masses. Our approved Cycle~6 program will complete our survey of $z \sim 4$ DLAs by including lower metallicity systems, and, thereby, establish the properties of the entire host population.

The large impact parameter ($\approx 16-45$~kpc) and high \HI\ column density along the DLA sightline suggest that a large fraction of the \HI\ resides in clumpy regions in the halo of high-$z$ galaxies, away from the bulk of the star formation. The excellent agreement between the absorption and emission redshifts for all five galaxies detected in \CII\ emission disfavors a scenario where the absorbing gas is solely in a companion galaxy as one would then expect velocity offsets between the centroids of the emission and absorption. The relatively high gas metallicity ($\sim 1/10^{\rm th}$ solar) along the DLA sightline indicates that metals must be effectively mixed with the gas and can escape from the main star-forming regions out to large distances. In summary, our ALMA observations suggest that the \HI\ distributions surrounding high-redshift galaxies are markedly different from the \HI\ distributions seen around galaxies in the local Universe.

Finally, we note that all quasars are detected in continuum emission, which is consistent with expectations based on the results of \citet{Decarli2018}. In addition, we have detected a bright continuum source close to one of our target quasars, QSO~J1253$+$1046. If the source is at the quasar redshift ($z \approx 4.908$), it would have a total infrared luminosity of $\approx 10^{13}$~L$_\odot$ and an SFR of $\approx 1000$~M$_\odot$~yr$^{-1}$, typical of high-$z$ sub-mm galaxies. This would be a second case of a quasar-SMG pair at these redshifts, after BR~1202$-$0725 at $z \approx 4.7$, but at an even smaller transverse separation, of only $\approx 12$~kpc.

\acknowledgments
We would like to thank the anonymous referee for helpful comments that have improved this manuscript. This paper makes use of the following ALMA data: ADS/JAO.ALMA \#2015.0.01564.S and \#2016.0.00569.S. ALMA is a partnership of ESO (representing its member states), NSF (USA) and NINS (Japan), together with NRC (Canada), MOST and ASIAA (Taiwan), and KASI (Republic of Korea), in cooperation with the Republic of Chile. The Joint ALMA Observatory is operated by ESO, AUI/NRAO and NAOJ. M.N. acknowledges support from ERC Advanced grant 740246 (Cosmic{\verb|_|}Gas). N.K. acknowledges support from the Department of Science and Technology via a Swarnajayanti Fellowship (DST/SJF/PSA-01/2012-13).

\begin{table*}[!b]
\centering
\caption{Properties of the DLAs and Associated ALMA \CII\ Emitters}
\label{tab:obs}
\begin{tabular}{cccccc}
 & QSO~J0834$+$2140 & QSO~J1101$+$0531 & QSO~J1253$+$1046 & \multicolumn{2}{c}{QSO~PSS1443$+$27}\\
\hline
\hline
\multicolumn{6}{c}{DLA Properties}\\
\hline
Right Ascension (J2000) & 08:34:29.44 & 11:01:34.36 & 12:53:53.35 & \multicolumn{2}{c}{14:43:31.17}\\
Declination (J2000) & $+$21:40:24.7 & $+$05:31:33.8 & $+$10:46:03.1 & \multicolumn{2}{c}{$+$27:24:36.7}\\
Redshift & 4.3900 & 4.3446 & 4.6001 & \multicolumn{2}{c}{4.2241}\\
$\log(N$(H\,\textsc{i})/cm$^{-2})$ & $21.00 \pm 0.20$ & $21.30 \pm 0.10$ & $20.30 \pm 0.15$ & \multicolumn{2}{c}{$21.00 \pm 0.10$}\\
$\rm{[M/H]}$ & $-1.30 \pm 0.20$ & $-1.07 \pm 0.12$ & $-1.36 \pm 0.16$ & \multicolumn{2}{c}{$-0.95 \pm 0.20$}\\
$\Delta V_{90}$ (\kms) & $290 \pm 10$ & $60 \pm 10$ & $70 \pm 10$ & \multicolumn{2}{c}{$284 \pm 10$}\\
\hline
\multicolumn{6}{c}{ALMA Observations}\\
\hline
Continuum Resolution ($'' \times ''$)$^{\rm a}$ & ($0.70 \times 0.56$) & ($0.99 \times 0.91$) & ($0.95 \times 0.83$) & \multicolumn{2}{c}{($0.67 \times 0.48$)}\\
Continuum RMS ($\mu$Jy beam$^{-1}$) & 16.4 & 24.5 & 22.7 & \multicolumn{2}{c}{26.2}\\
Cube Resolution ($'' \times ''$)$^{\rm a}$ & ($0.69 \times 0.56$) & ($0.97 \times 0.89$) & ($0.96 \times 0.82$) & \multicolumn{2}{c}{($0.67 \times 0.47$)}\\
Channel width (km~s$^{-1}$) & 53.1 & 52.7 & 55.2 & \multicolumn{2}{c}{51.5}\\
RMS per channel (mJy beam$^{-1}$) & 0.17 & 0.27 & 0.19 & \multicolumn{2}{c}{0.29}\\
\hline
\multicolumn{6}{c}{\CII\ Emitter Properties}\\
\hline
Right Ascension (J2000) & 08:34:29.71 &  11:01:34.34 & --- & 14:43:31.29 & 14:43:31.41\\
Declination (J2000) & $+$21:40:23.3 & $+$05:31:37.8 & --- & $+$27:24:38.3 & $+$27:24:38.8\\
Redshift & 4.3896 & 4.3433 & --- & 4.2227 & 4.2276\\
Impact Parameter ($''$) & 4.0 & 4.0 & --- & 2.3 & 4.0\\
$S_{\rm cont}$ ($\mu$Jy) & $70 \pm 22$ & $<75$$^{\rm b}$ & $<84$$^{\rm b,c}$ & $161 \pm 44$ & $<81$$^{\rm b}$\\
$\int S_{\rm [C{\scriptscriptstyle II}]}dV$ (Jy km s$^{-1}$) & $0.173 \pm 0.020$ & $0.062 \pm 0.012$ & $<0.042$$^{\rm b,c}$ & $0.846 \pm 0.063$ & $0.274 \pm 0.043$\\
$\rm{[C}$\,\textsc{ii}]-$\Delta V_{90}$ (km s$^{-1}$)$^{\rm d}$ & $270 \pm 60$ & $370 \pm 60$ & --- & $510 \pm 60$ &  $620 \pm 60$\\
\LCII\ ($10^{8}$ L$_\odot$) & $1.02 \pm 0.12$ & $0.36 \pm 0.07$ & $<0.27$$^{\rm b,c}$ & $4.7 \pm 0.4$ & $1.5 \pm 0.2$\\
$L_{\rm TIR}$ ($10^{10}$ L$_\odot$)$^{\rm e}$ & $6.6 \pm 2.1$ & $< 7.0$$^{\rm b}$ & $<8.5$$^{\rm b,c}$ & $14 \pm 4$ & $<7.2$$^{\rm b}$\\
SFR (M$_\odot$ yr$^{-1}$) & $7 \pm 2$ & $<7$$^{\rm b}$ & $<9$$^{\rm b,c}$ & $15 \pm 4$ & $<7$$^{\rm b}$\\

\hline
\multicolumn{6}{l}{$^{\rm a}$ The FWHM of the synthesized beam of the image.}\\
\multicolumn{6}{l}{$^{\rm b}$ All upper limits are 3$\sigma$ and assuming the source is unresolved in the observations.}\\
\multicolumn{6}{l}{$^{\rm c}$ Limit for a source within 10$''$ of the DLA sight line, and assuming a FWHM line width of 100 \kms.}\\
\multicolumn{6}{l}{$^{\rm d}$ Velocity interval that contains 90 \% of the velocity-integrated \CII\ flux density.}\\
\multicolumn{6}{l}{$^{\rm e}$ Systematic uncertainties are 0.5 dex.}
\end{tabular}
\end{table*}

\end{document}